# Trust Management in Cloud Computing: A Critical Review

Mohamed Firdhous, Osman Ghazali and Suhaidi Hassan

*Abstract*—Cloud computing has been attracting the attention of several researchers both in the academia and the industry as it provides many opportunities for organizations by offering a range of computing services. For cloud computing to become widely adopted by both the enterprises and individuals, several issues have to be solved. A key issue that needs special attention is security of clouds, and trust management is an important component of cloud security.

In this paper, the authors look at what trust is and how trust has been applied in distributed computing. Trust models proposed for various distributed system has then been summarized. The trust management systems proposed for cloud computing have been investigated with special emphasis on their capability, applicability in practical heterogonous cloud environment and implementabilty. Finally, the proposed models/systems have been compared with each other based on a selected set of cloud computing parameters in a table.

*Index Terms*—Cloud Computing, Trust, Trust Management, Trust Models

## I. INTRODUCTION

Distributed systems like peer-to-peer systems, grid, clusters and cloud computing have become very popular among users in the recent years. Users access distributed systems for different reasons such as downloading files, searching for information, purchasing goods and services or executing applications hosted remotely. With the popularity and growth of distributed systems, service providers make new services available on the system. All these services and service providers will have varying levels of quality and also, due to the anonymous nature of the systems, some unscrupulous providers may tend to cheat unsuspecting clients. Hence it becomes necessary to identify the quality of services and service providers who would meet the requirements of the customers [1].

Mohamed Firdhous is a Senior Lecturer attached to the Faculty of Information Technology, University of Moratwua, Sri Lanka. (email: firdhous@itfac.mrt.ac.lk & mfirdhous@internetworks.my)

Dr. Osman Ghazali is a Senior Lecturer and the Head of School of Computing, College of Arts and Sciences of the Universiti Utara Malaysia. (email: osman@uum.edu.my)

Prof. Suhaidi Hassan is an Associate Professor and the Asst. Vice Chancellor of the College of Arts and Sciences of the Universiti Utara Malaysia. He is the team leader of the InterNetWorks Research Group. (email: suhaidi@uum.edu.my)

In this paper the authors take a look at the trust and trust management systems along with the trust models developed for distributed systems. Then a critical look at the trust development and management systems for cloud computing systems reported in literature in the recent times has been taken with special reference to the pros and cons of each proposal.

## II. CLOUD COMPUTING

Cloud computing has been called the $5^{th}$ utility in line of electricity, water, telephony and gas [2]. The reason why cloud has been nomenclature with such a name is that cloud computing has been changing the way computer resources have been used up to now. Until the development of cloud computing, computing resources were purchased outright or leased in the form of dedicated hardware and software resources. Cloud computing has brought a paradigm change in how computing resources have been purchased. With the advent of cloud computing, users can use the services that have been hosted on the internet without worrying about whether they have been hosted or managed in such a manner that the customers have to pay only for the services they consumed as in the case of making use of other services.

Cloud providers host their resources on the internet on virtual computers and make them available to multiple clients. Multiple virtual computers can run on one physical computer sharing the resources such as storage, memory, the CPU and interfaces giving the feeling to the client that each client has his own dedicated hardware to work on. Virtualization thus gives the ability to the providers to sell the same hardware resources among multiple clients. This sharing of the hardware resources by multiple clients help reduce the cost of hardware for clients while increasing profits of providers. Accessing or selling hardware in the form of virtual computers is known as Infrastructure as Service (IaaS) in the cloud computing terminology [3]. Once a client has procured infrastructure from a service provider, he is free to install and run any Operating System platform and application on it.

Other kinds of services that are made available via the cloud computing model are Platform as a Service (PaaS) and Software as a Service. Figure 1, shows the architecture of a typical cloud computing system.

Under PaaS, the development platform in the form of an Operating System has been made available where customers can configure the environment to suit their requirements and install their development tools [5]. PaaS helps developers develop and deploy applications without the cost of purchasing and managing the underlying hardware and software. PaaS provides all the required facilities for the complete life cycle of building and delivering web applications. Thus PaaS usually offers facilities for application design, application development, testing, deployment and hosting as well as application services such as team collaboration, web service integration and marshalling, database integration, security, scalability, storage, persistence, state management, application versioning, application instrumentation and developer community facilitation.

SaaS is the cloud model where an application hosted by a service provider on the internet is made available to users in a ready to use state. SasS eliminates the requirement of installation and maintenance of the application in the user's local computer or server in his premises [5]. SaaS has the advantage of being accessible from any place at any time, no installation or maintenance, no upfront cost, no licensing cost, scalability, reliability and flexible payment schemes to suit the customer's requirements.

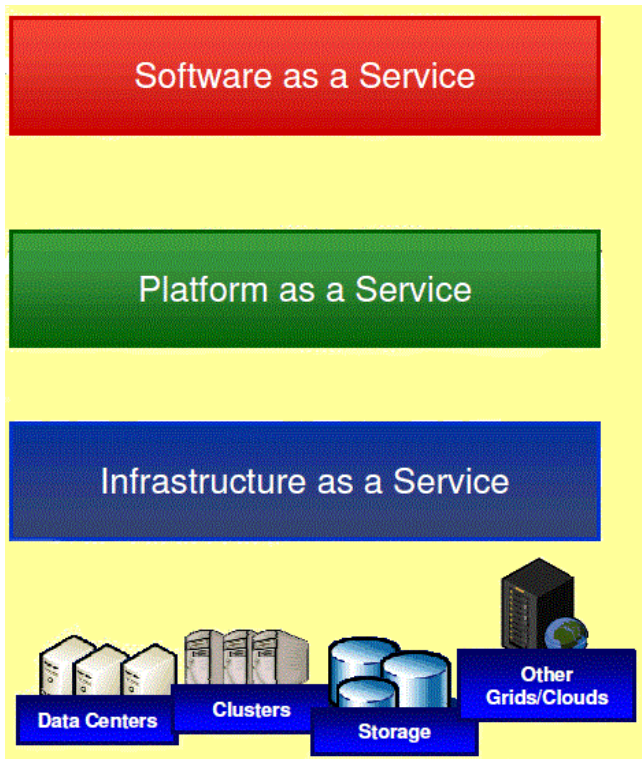

Fig. 1. Cloud Computing Architecture

### III. TRUST AND TRUST MANAGEMENT

The trust and reputation have their origin in the social sciences that study the nature and behavior of human societies [6]. Trust has been studied by researchers in diverse fields such as psychology, sociology and economics [7]. Psychologists study trust as a mental attitude and focus on what happens in a person's mind when he/she trusts or distrusts someone [8]. Based on this notion, several cognitive trust models have been developed [9-12]. Sociologists approach to trust as a social relationship between people. Social context of trust has been commonly employed in multi agent systems and social networks [7,13-14]. The similarity between multi agent system and a social network are exploited in these works as agents and people behave in a similar fashion interacting with, gathering information from and modeling each other for developing trust in each other. Economists perceive trust in terms of utility [15]. Game theory has been one of the most popular tools used by experts in the computer field to study how users develop trust using different strategies [16-17]. The prisoner's dilemma is the commonly used scenario to study this scenario [18-19].

Researchers in computer sciences have exploited the benefit of all these studies as they provide vital insight into human behavior under various circumstances [13, 20-21]. The role of trust and reputation in open, public distributed systems such as e-commerce, peer to peer networks, grid computing, semantic web, web services and mobile networks have been studied by several researchers [22-25].

Although the rich literature available on trust from diverse fields is of great benefit to computer scientists, it has the drawback of presenting a complex and confusing notion for trust. This is mainly due to the reason that there is no common agreement of a single definition for what trust is? It can be seen that different researchers have defined trust as attitudes, beliefs, probabilities, expectations, honesty and so on.

Even if different disciplines and researchers look at trust from different angles, it is possible to identify some key factors that are common to everything. They are;

- Trust plays a role only when the environment is uncertain and risky.
- Trust is the basis based on which certain decisions are made.
- Trust is built using prior knowledge and experience.
- Trust is a subjective notion based on opinion and values of an individual.
- Trust changes with time and new knowledge while experience will have overriding influence over the old ones.
- Trust is context-dependent.
- Trust is multi-faceted.

McKnight and Chervany have identified 16 characteristics of trust and grouped them under five groups. They are,

- Competence; competent, expert, dynamic
- Predictability; predictable
- Benevolence; good (or moral), good-will benevolent (caring), responsive
- Integrity; honest, credible, reliable, dependable
- Other; open, careful (or safe), shared understanding, personally attractive [8].

De Oliveira and Maziero have classified trust relations into hierarchical trust, social groups and social networks.

Hierarchical trust considers all relationships in a hierarchical manner and represented by a tree organization where nodes represent individuals and edges represent the trust degrees between the pair of nodes. Any two nodes can define a trust degree between them through transitivity through other nodes [26].

Zhang et al., have classified the trust functions based on the following four dimensions [27].

- Subjective trust vs. Objective trust
- Transaction-based vs. Opinion-based
- Complete information vs. Localized information
- Rank-based vs. Threshold-based

Capability of an entity's trustworthiness being measured objectively against a universal standard, results in objective trust. If the trust being measured depends on an individual's tastes and interest, the resulting trust is called *subjective trust*. Decisions made based on the individual transactions and their results is known as *transaction based trust*, whereas the trust built based on just opinion of the individuals, is *opinion based trust*. If the trust building operation requires information from each and every node, it is called, complete information and it is known as either *global trust function* or *complete trust function*. If the information collected only from one's neighbors, it is called, *localized information trust function*. If the trust worthiness of an entity is ranked from the best to worst, it is *rank based trust* whereas the trust declared yes or no depending on? Preset trust threshold is known as *threshold based trust*.

## IV. TRUST MODELS

Several models have been developed by researchers for the purpose of building practical trust systems in distributed systems. This section takes a brief look at some of the commonly used trust models.

### A. CuboidTrust

CuboidTrust is a global reputation-based trust model for peer to peer networks. It takes three factors namely, contribution of the peer to the system, peer's trustworthiness in giving feedback and quality of resources to build four relations. Then it creates a cuboid using small cubes whose coordinates (x,y,z) where z – quality of resource, y – peer that stores the value and x – the peer which rated the resource and denoted by Px,y,z. The rating is binary, 1 indicating authentic and (–1) indicating inauthentic or no rating. Global trust for each peer has been computed using power iteration of all the values stored by the peers [28].

### B. EigenTrust

EigenTrust assigns each peer a unique global trust value in a P2P file sharing network, based on the peer's history of uploads. This helps to decrease the downloading of inauthentic files. Local trust value $S_{ij}$ has been defined $S_{ij} = sat(i,j) - unsat(i,j)$, where $sat(i,j)$ denotes the satisfactory downloads by $i$ from $j$ and $unsat(i,j)$ is the unsatisfactory downloads by $i$ from $j$. Power iteration is used to compute the global trust for each peer [29].

### C. Bayesian Network based Trust Management (BNBTM)

BNBTM uses multidimensional application specific trust values and each dimension is evaluated using a single Bayesian network. The distribution of trust values is represented by beta probability distribution functions based on the interaction history [30].

Trust value of peer $i$ is given by,

$$\tau_i = \frac{\alpha_i}{\alpha_i + \beta_{\bar{\imath}}}, (i \in \{G, L, C\}, \bar{\imath} \in \{\bar{G}, \bar{L}, \bar{C}\}) \quad (1)$$

Where $\alpha_i = r_i + 1$ and $\beta_{\bar{\imath}} = s_{\bar{\imath}} + 1$ and $r_i$ and $s_{\bar{\imath}}$ are number of interactions with outcome $i$ and $\bar{\imath}$ respectively. $G, L$ and $C$ represent shipping goods, shipping lower quality goods and not shipping any goods and $\bar{G}, \bar{L}$ and $\bar{C}$ represent the converse.

### D. GroupRep

GroupRep is a group based trust management system. This classifies trust relationships in three levels namely, trust relationships between groups, between groups and peers and only between peers [31].

Trust of Group $i$ held by Group $j$ is given by:

$$T_{rG_iG_j} = \begin{cases} \frac{u_{G_iG_j} - c_{G_iG_j}}{u_{G_iG_j} + c_{G_iG_j}} & \text{if } u_{G_iG_j} + c_{G_iG_j} \neq 0 \\ T_{rG_iG_j}^{reference} & \text{if } u_{G_iG_j} + c_{G_iG_j} = 0 \text{ and } \exists Trust_{G_iG_j}^{path} \\ T_{rG_iG_{strange}} & \text{otherwise} \end{cases} \quad (2)$$

Where $u_{G_iG_j} \geq 0$ and $c_{G_iG_j} \geq 0$ are utility and cost respectively assigned by nodes in group $j$ to nodes in group $i$.

$T_{rG_iG_j}^{reference}$ is defined as the minimum trust value along the most trustworthy reference path.

### E. AntRep

AntRep algorithm is based on swarm intelligence. In this algorithm, every peer maintains a reputation table similar to distance vector routing table. The reputation table slightly differs from the routing table in the sense that (i) each peer in the reputation table corresponds to one reputation content; (ii) the metric is the probability of choosing each neighbor as the next hop whereas in the routing table it is the hop count to destinations. Both forward ants and backward ants are used for finding reputation values and propagating them. If the reputation table has a neighbor with the highest reputation, a unicast ant is sent in that direction. If no preference exists, broadcast ants are sent along all the paths [32].

Once the required reputation information is found, a backward ant is generated. When this ant travels back, it updates all the reputation tables in each node on its way.

## F. Semantic Web

Zhang et al., have presented a trust model which searches all the paths that connect the two agents to compute the trustworthiness between those two agents. For each path the ratings associated with each edge are multiplied and finally all the paths are added to calculate the final trust value [33].

The weight of the path $i$ ($w_i$) is calculated using;

$$w_i = \frac{\frac{1}{D_i}}{\sum_{i=1}^{N} \frac{1}{D_i}} \quad (3)$$

Where  $N$ – No. of paths between agents $P$ and $Q$
$D_i$ – No. of steps between $P$ and $Q$ on the $i^{th}$ path.
$m_i$ – Q's immediate friend or neighbor on the $i^{th}$ path. (M – set of Q's friends or neighbors)

This gives a higher weight to shorter paths.

If agent $P$ and agent $Q$ are friends then $P \rightarrow Q$, or neighbors then $P \leftrightarrow Q$ then $P$'s trust in $Q$ can be computed directly. Otherwise,

$$T_{P \rightarrow Q} = \sum_{i=1}^{N} \frac{T_{m_i \rightarrow Q} \times \prod_{i \rightarrow j \cup i \leftrightarrow j} R_{i \rightarrow j} \times \frac{1}{D_i}}{\sum_{i=1}^{N} \frac{1}{D_i}}$$

$$= \sum_{i=1}^{N} T_{m_i \rightarrow Q} \times \prod_{i \rightarrow j \cup i \leftrightarrow j} R_{i \rightarrow j} \times w_i \quad (4)$$

Where reliability factor $R_{i \rightarrow j}$ denotes to which degree $i$ believes in $j$'s words or opinions.

## G. Global Trust

Several authors have presented methods that compute an improved global trust value for selecting trusted source peer in peer to peer systems [34-36].

The global trust value for node $i$, $t_i$ is defined as:

$$t_i = \sum_k c_{ki} t_k \quad (5)$$

Where $c_{ki}$ is the local trust value from peer $k$ towards peer $i$ and $t_k$ is the global trust value of peer $k$.

## H. Peer Trust

This is reputation-based trust supporting framework. This includes a coherent adaptive trust model for quantifying and comparing the trustworthiness of peers based on a transaction-based feedback system. It introduces three basic trust parameters namely feedback a peer receives from other peers, the total number of transactions a peer performs, the credibility of the feedback sources and two adaptive factors that are transaction context factor and the community context factor in computing trustworthiness of peers, then it combines these factors to compute a general trust metric [37].

## I. PATROL-F

PATROL-F incorporates many important concepts for the purpose of computing peer reputation. The main components used in computing peer trust are: direct experiences and reputation values, the node credibility to give recommendations, the decay of information with time based on a decay factor, first impressions and a node system hierarchy [38].

It uses three fuzzy subsystems:

1. The first is used to set the importance factor of an interaction and related decisions. To decide and choose which data is critical or indispensable, or which data is needed more quickly, is a concept close to humans that fuzzy logic can model.

2. Then there is the region of uncertainty where an entity is not sure whether to trust or not (when the reputation of a host between the absolute mistrust level φ, and the absolute trust level θ ). Fuzzy techniques are effectively applied in this region.

3. Finally, for the Result of Interaction (RI) value, fuzzy logic can be used to capture the subjective and humanistic concept of four level *"good"* or *"better"* and *"bad"* or *"worse"* interaction. RI is the result of several concepts effectively combined to produce a more representative value. The decay factor τ is calculated based on the difference of a host's values of RIs between successive interactions.

## J. Trust Evolution

Wang et al., have presented a trust evolution model for P2P networks. This model uses two critical dimensions, experience and context to build trust relationships among peers. It builds two kinds of trust: direct trust and recommendation trust quantifies trust within the interval *[0,1]* [39].

Direct trust (DT) between two peers is computed using the last *n* interactions between those entities. Recommended trust is calculated using recommendations from other peers and the previous interactions with the recommending peers.

## K. Time-based Dynamic Trust Model (TDTM)

TDTM is an ant colony based system that identifies the pheromone and the trust and the heuristic and the distance between two nodes. The trust value calculated by this model depends on the frequency of interaction where the trust value increases with frequent interactions and lowers as the interactions goes down [40].

Trust-pheromone between nodes *i* and *j* at time (*t +1*) is defined as:

$$\tau_{ij}(t+1) = \rho \tau_{ij}(t) + \sigma \tau_{ij}(t) \quad (6)$$

Where $\rho$ is the trust dilution factor and $\sigma\tau_{ij}(t)$ is the additional intensity at each inter-operation between entities.

$\sigma\tau_{ij}(t)$ is defined as:

$$\sigma\tau_{ij} = \begin{cases} \frac{1}{\frac{1}{1-\tau_{ij}(t)}+1} & \text{if } i \text{ and } j \text{ interact at time } t \\ 0 & \text{otherwise} \end{cases} \quad (7)$$

If the trust value $p_{ij}(t)$ between nodes $i$ and $j$ at time $t$ is greater than a certain threshold $R$, they can validate each other's certificate, otherwise not.

*L. Trust Ant Colony System (TACS)*

TACS is based on the bio-inspired algorithm of ant colony system. In this model pheromone traces are identified with the amount of trust a peer has on its neighbors when supplying a specific service. It computes and selects both the most trustworthy node to interact and the most trustworthy path leading to that peer. Each peer needs to keep track of the current topology of the network as every peer has its own pheromone traces for every link. Ants travel along every path searching building the most trustworthy path leading to the most reputable server [41].

Ants stop the search once they find a node that offers the service requested by the client and the pheromone traces belonging to the current path leading to it are above the preset threshold, otherwise they would follow on further selecting a neighbor that has not been visited yet.

*M. TRUMMAR (TRUst Model for Mobile Agent systems based on Reputation)*

TRUMMAR is a general model for the calculation of reputation values and the determination of trust decisions. TRUMMAR identifies three types of nodes from who it can receive trust values. They are neighbors, friends and strangers. Neighbors are the trusting other hosts on its own network that are under the same administrative control, friends are the hosts from different networks that are under different, but trusted administrative control and strangers are the hosts that are willing to volunteer information but not neighbors or friends [42].

The trust value for Y in X is calculated as follows:

$$\frac{repY}{X(0)} = A\frac{repY}{X} + B\frac{\sum_i \frac{\alpha_i repY}{X_i}}{\sum_i \alpha_i} + C\frac{\sum_j \frac{\beta_j repY}{X_j}}{\sum_j \beta_j} + D\frac{\sum_l \frac{\delta_l repY}{X_l}}{\sum_l \delta_l} \quad (8)$$

Where

$\frac{repY}{X(0)}$ represents the reputation value being calculated.

$\frac{repY}{X(0)}$ represents the reputation value last calculated, modified to account for the time lapsed.

$\frac{\sum_i \frac{\alpha_i repY}{X_i}}{\sum_i \alpha_i}$ weighted sum of reputation reported by neighbors.

$\frac{\sum_j \frac{\beta_j repY}{X_j}}{\sum_j \beta_j}$ weighted sum of reputation reported by friends.

$\frac{\sum_l \frac{\delta_l repY}{X_l}}{\sum_l \delta_l}$ weighted sum of reputation reported by strangers.

$\alpha_i$, $\beta_j$ and $\delta_l$ are weighing factors which depend on the reputation of the individual neighbors, friends, and strangers in the host space, respectively.

A, B, C, and D are weighing factors for the respective reputation of with respect to self, neighbors, friends and strangers in the agent space and A > B > C > D.

Reputation values are restricted to values between 0 and k, i.e $0 \leq \frac{repY}{X}$;

*N. PATROL (comPrehensive reputAtion-based TRust mOdeL)*

PATROL is a general purpose reputation based trust model for distributed computing. PATROL is an enhancement over TRUMMAR. This model is based on multiple factors such as reputation values, direct experiences, trust in the recommender, time dependence of the trust value, first impressions, similarity, popularity, activity, cooperation between hosts, and hierarchy of host systems. The decision to interact with another host depends on two factors namely, the trust in the competence of a host and the trust in the host's credibility to give trusted advice. The trust in the competence of a host is calculated from the direct interactions and this is the confidence that the other host would be able to complete the intended task to the initiator host's expectations. The trust in a host's ability to give trusted advice is the confidence that the host gives consistent and credible advice and feedback. The overall trust value is a combination of the weighted values calculated for different factors calculated independently [43].

The operation of the model is as given below:

1. Host X wants to interact with host Y.

2. X calculates the time since it interacted last with Y, if this time is smaller than a predetermined threshold, it will decay the stored trust value compare against a predetermined threshold. If larger than the threshold, it will interact with Y, otherwise not.

3. If the last interaction time was larger than the threshold, it will involve other trusted hosts in its calculation of trust value for Y. If not,

4. Queried hosts will decay their stored trust value for Y and send it along with their reputation vectors.

5. X will calculate the trust for Y and check against the threshold. If the trust value is greater than the threshold, it will interact with Y, otherwise no interaction.

### O. META-TACS

META-TACS is an extension of the TACS algorithm developed by the [41]. They have extended the TACS model by optimizing the working parameters of the algorithm using genetic algorithms [44].

### P. CATRAC (Context-Aware Trust- and Role-Based Access Control for composite web services)

Role-Based Access Control (RBAC) and Trust-Based Access Control (TBAC) have been proposed to address threats to security in single Web Service scenarios. But these solutions fail to provide the required security level in situations related to composite Web Services. CATRAC has been proposed as a security framework related to composite web services [45]. CATRAC combines both RBAC and TBAC in order to arrive at an optimum solution.

Three conditions must be satisfied to gain access to a specific web service. They are:

- Client attributes must be authenticated by the web service provider.
- Client's global role must be valid and contains the right permissions.
- Client's trust level must be equal or greater than the threshold level set for the particular service.

A trusted third party called the Role Authority issues, signs and verifies the roles assigned to the clients. Trust levels are expressed as a vector ranging from 0 to 10, indicating the fully distrusted to the fully trusted respectively. Five (5) indicates a neutral or uncertainty level which is commonly assigned to new clients.

CATRAC is made up of three entities, namely Role Authority, Servers and Clients. Clients accumulate trust points when their behavior is considered good and otherwise they lose trust points. Also, clients trust level is decayed to the neutral value gradually with time, if no interaction takes place. Trust level is decayed using the following formulae.

$$D_{TL_c} = \left( (TL_c - TL_N) \times e^{\left(\frac{-t}{memo_s}\right)} \right) + TL_N \quad (9)$$

If the current trust level is above the neutral trust level.

$$D_{TL_c} = TL_N \times e^{\left(\frac{-t}{memo_s}\right)} \quad \text{otherwise.} \quad (10)$$

Where

$D_{TL_c}$ – decayed trust level for client c
$TL_c$ – current trust level for client c
$TL_N$ – neutral trust level
$t$ – time elapsed
$memo_s$ – memory factor (constant)

### Q. Bayesian Network -based Trust Model

Bayesian Network–based Trust Model computes trust values by combining multiple input attributes [46]. In this model, the different capabilities of providers such as the type of the file, quality of the file, download speed etc. Also, it looks at the contextual representation of trust values. That is, if two agents compute the trust values, they can trust each other's recommendation and if the agents use different criteria, they may not trust the each other's recommendation even if both are truthful.

In this system each peer identified as an agent develops a naïve Bayesian network for each provider it has interacted with. Each Bayesian network has a root node T with two branches named "satisfying" and "unsatisfying", denoted by 1 and 0, respectively. The agents overall trust in the provider's competencies represented by *p(T=1)*, which is the ratio of interactions with satisfactory results out of all the interactions with the same provider. On the other hand *p(T=0)* is the ratio of unsatisfactory results under the same criteria.

Hence: *p(T=1) + p(T=0) = 1*         (11)

Depending on the results of the previous interactions, the agent creates a conditional probability in the form of *p(File Type = "Music" | T = 1)* or *p(Download Speed = "High" | T = 1)* for each quality attribute such as file type, file quality and speed. These conditional probability values are stored in a table called the Conditional Probability Table (CPT).

Finally the provider's trustworthiness in different aspects such as *p(T = 1 | File Type ="Music" AND Download Speed = "High")* is computed by combing the conditional probability values stored in the CPT using the Bayes rule. This combined trustworthiness value is the overall trust score of the provider for the given attribute(s) or aspect(s).

The models discussed above have been proposed for different types of distributed systems such as clusters, grids and wireless sensor networks. But none of the above models has been tested on the cloud computing environment. Hence an extensive evaluation of these models needs to be carried out to understand the advantages and disadvantages of these models for use in cloud computing. The authors propose to carry out this kind of evaluation of these models in future work. Next section takes an in depth look at the trust models proposed for cloud computing.

## V. TRUST IN CLOUD COMPUTING

Security is one of the most important areas to be handled in the emerging area of cloud computing. If the security is not handled properly, the entire area of cloud computing would fail as cloud computing mainly involves managing personal sensitive information in a public network. Also, security from the service providers point also becomes imperative in order to protect the network, the resources in order to improve the robustness and reliability of those resources. Trust management that models the trust on the behavior of the elements and entities would be especially useful for the proper administration of cloud system and cloud services.

Several leading research groups both in academia and the industry are working in the area of trust management in cloud computing. This section takes an in depth look at the recent developments in this area with the objective of identifying and categorizing them for easy reference.

Khan and Malluhi have looked at the trust in the cloud system from a users perspective. They analyze the issues of trust from what a cloud user would expect with respect to their data in terms of security and privacy. They further discuss that what kind of strategy the service providers may undertake to enhance the trust of the user in cloud services and providers. They have identified control, ownership, prevention and security as the key aspects that decide users' level of trust on services. Diminishing control and lack of transparency have identified as the issues that diminishes the user trust on cloud systems. The authors have predicted that remote access control facilities for resources of the users, transparency with respect to cloud providers actions in the form of automatic traceability facilities, certification of cloud security properties and capabilities through an independent certification authority and providing security enclave for users could be used to enhance the trust of users in the services and service providers [47].

Zhexuan et al., have taken a look at the security issues SaaS might create due to the unrestricted access on user data given to the remotely installed software [48]. The authors have presented a mechanism to separate software from data so that it is possible to create a trusted binding between them. The mechanism introduced involves four parties namely the resource provider, software provider, data provider and the coordinator. The resource provider hosts both data and software and provides the platform to execute the software on data. The software provider and data provider are the owners of the software and data respectively. The coordinator brings the other parties together while providing the ancillary services such as searching for resources and providing an interface to execute the application on the data.

The operation of the model is as follows:

Software provider and data provider upload their resources to the resource provider. These resources will be encrypted before stored and the key will be stored in the accountability vault module of the system.

A data provider searches for and finds the required software through a coordinator and then runs the software on the data uploaded to the resource provider's site.

Once the execution has started an execution reference ID is generated and given to the data provider.

When the execution of the software is over, the results are produced only on the data provider's interface which can be viewed, printed or downloaded.

Data provider will then pay for the service that will be split between the software provider and resource provider.

An operation log has been created and posted to the software provider without disclosing the data provider's identity or the content on which software was run. This helps the software provider know that his software has been used and the duration of use.

Even though the authors claim that this model separates the software and data, there is no assurance that the software cannot make a copy while the data is being processed as only the algorithm or description of the software is provided to the data owner. Without the source code, there is no assurance that the code will not contain any malicious code hidden inside. Also, since the software runs on data owner's rights and privileges, the software would have complete control over data. This is a security threat and the audit trail even if it is available, will not detect any security breaches.

The authors do not address the question of trust on the proposed platform as this would be another application or service hosted on the cloud. Both application providers and data providers need some kind of better assurance as now they are entrusting their data and software to a third party software.

Sato et al., have proposed a trust model of cloud security in terms of social security [49]. The authors have identified and named the specific security issue as social insecurity problem and tried to handle it using a three pronged approach. They have subdivided the social insecurity problem in to three sub areas, namely; multiple stakeholder problem, open space security problem and mission critical data handling problem.

The multiple stakeholder problem addresses the security issues created due to the multiple parties interacting in the cloud system. As per the authors, three parties can be clearly identified. They are namely, the client, the cloud service providers and third parties that include rivals and stakeholders in business. The client delegates some of the administration/operations to cloud providers under a Service Level Agreement (SLA). Even if the client would like to have the same type of policies that it would apply if the resources were hosted on site on the delegated resources, the provider's policy may differ from that of the client. The providers are bound only by the SLA signed between the parties. The SLA plays the role of glue between the policies. Also the authors opine that once the data is put in the cloud it is open for access by third parties once authenticated by the cloud provider.

The open space security problem addresses the issue of loss of control on where the data is stored and how they are physically managed once control of data is delegated to the cloud provider. They advice to encrypt the data before transferring, converting the data security problem to a key management problem as now the keys used for encryption/decryption must be handled properly.

The mission critical data handling problem looks at the issue of delegating the control of mission critical data to a service provider. They advice not to delegate control of this data but to keep them in a private cloud in a hybrid setup, where the organization have unhindered control. However setting up of a private cloud may not be an option to small and medium sized organizations due to the high costs involved. Hence enhancement of security of the public cloud is the only option to serve everybody.

Authors have developed a trust model named 'cloud trust model' to address the problems raised above. Two more trust layers have been added to the conventional trust architecture. These layers have been named as The internal trust layer and the contracted trust layer. The Internal trust layer acts as the platform to build the entire trust architecture. It is installed in the in house facilities and hence under the control of the local administration. ID and key management are handled under the internal trust. Also any data that is considered critical or needs extra security must be stored under this layer.

Contracted trust has been defined as the trust enforced by an agreement. A cloud provider places his trust upon the client, based on the contract that is made up of three documents known as, Service Policy/Service Practice Statement (SP/SPS), Id Policy/Id Practice Statement (IdP/IdPS) and the contract.

Level of trust required can be negotiated by parties depending on the level of security needed for the data. A cloud system thus installed is called a secure cloud by the authors.

Li et al., propose a domain-based trust model to ensure the security and interoperability of cloud and cross-clouds environment and a security framework with an independent trust management module on top of traditional security modules [50]. They also put forward some trust based security strategies for the safety of both cloud customers and providers based on this security model.

A cloud trust model based on the family gene technology that is fundamentally different from the Public key Infrastructure based trust models has been proposed by Wang et al.,. The authors have studied the basic operations such as user authentication, authorization management and access control and proposed a Family-gene Based model for Cloud Trust (FBCT) integrating these operations [51-52].

Manuel et al., have proposed trust model that is integrated with CARE resource broker [53]. The proposed trust model can support both grid and cloud systems. The model computes trust using three main components namely, Security Level Evaluator, Feedback Evaluator and Reputation Trust Evaluator. Security Level Evaluation has been carried out based on authentication type, authorization type and self security competence mechanism. Multiple authentication, authorization mechanism and self security competence mechanisms are supported. Depending on the strength of individual mechanism, different grades are provided for trust value. Feedback Evaluation also goes through three different stages namely feedback collection, feedback verification and feedback updating. The Reputation Trust Evaluator computes the trust values of the grid/cloud resources based on their capabilities based on computational parameters and network parameters. Finally the overall trust value has been computed taking the arithmetic sum of all the individual trust values computed.

Shen et al., and Shen and Tong have analyzed the security of cloud computing environment and described the function of trusted computing platform in cloud computing [54-55]. They have also proposed a method to improve the security and dependability of cloud computing integrating the Trusted Computing Platform (TCP) into the cloud computing system. The TCP has been used in authentication, confidentiality and integrity in cloud computing environment. Finally the model has been developed as software middleware known as the Trusted Platform Software Stack (TSS).

Alhamad et al., have proposed a SLA based trust model for cloud computing. The model consists of the SLA agents, cloud consumer module and cloud services directory [56]. The SLA agent is the core module of the architecture as it groups the consumers to classes based on their needs, designs SLA metrics, negotiates with cloud providers, selects the providers based on non functional requirements such as QoS, and monitors the activities for the consumers and the SLA parameters. Cloud consumer module requests the external execution of one or more services. Cloud services directory is the one where the service providers can advertise their services and consumers seek to find the providers who meet their functional requirements such as database providers, hardware providers, application providers etc.,

The authors have proposed only the model and no implementation or evaluation has been developed or described. Hence the each and every module will have to be evaluated for their functionality and the effectiveness and finally the overall model will have to be evaluated for its effectiveness.

Yong et al., have proposed a model called a multi-tenancy trusted computing environment model (MTCEM) for cloud computing [57]. MTCEM has been proposed to deliver trusted IaaS to customers with a dual level transitive trust mechanism that supports a security duty separation function simultaneously. Since cloud facilities belong to multiple stakeholders such as Cloud Service Providers (CSP) and customers, they belong to multiple security domain and server different security subjects simultaneously. The different stakeholders may be driven by different motives such as best service, maximization of the return on investment and hence may work detrimental to the other party involved. Hence cloud computing should have the capability to compartmentalize each customer and CSP and support security duty separation defining clear and seamless security responsibility boundaries for CSP and customers.

MTCEM has been designed as two-level hierarchy transitive trust chain model which supports the security duty separation and supports three types of distinct stakeholders namely, CSP, customers and auditors. In this model, CSP assume the responsibilities to keep infrastructures trusted while the customer assumes responsibility starting from the guest OS which installed by the customer on the Virtual Machines provided by the CSP. The auditor monitors the services provided by the CSP on behalf of the customers. The authors have implemented a prototype system to prove that MTCEM is capable of being implemented on commercial hardware and software. But no evaluation of the prototype on performance has been presented.

Yang et al., have studied the existing trust models and firewall technology. The authors have found that all the existing trust models ignore the existence of firewall in a network [58]. Since firewall is an integral and important component of any corporate security architecture, this non inclusion of firewall is a huge shortcoming. The authors have proposed a collaborative trust model of firewall-through based on Cloud theory. This paper also presents the detailed design calculations of the proposed trust model and practical algorithms of measuring and updating the value of dynamic trust.

The model has the following advantages compared to other models:

- There are different security policies for different domains.
- The model considers the transaction context, the historical data of entity influences and the measurement of trust value dynamically.
- The trust model is compatible with the firewall and does not break the firewall's local control policies.

Fu et al., have studied the security issues associated with software running in the cloud and proposed a watermark-aware trusted running environment to protect the software running in the cloud [59]. The proposed model is made up of two components namely the administrative center and the cloud server environment. The administrative center embeds watermark and customizes the Java Virtual Machines (JVM) and the specific trusted server platform includes a series of cloud servers deployed with the customized JVMs. Only specific and complete Java programs are allowed to run on the JVMs while rejecting all the unauthorized programs like invasion programs. The main advantage of this approach is that it introduces watermark aware running environment to cloud computing.

Ranchal et al., have studied the identity management in cloud computing and proposed a system without the involvement of a trusted third party [60]. The proposed system that is based on the use of predicates over encrypted data and multi-party computing is not only capable of using trusted hosts but also untrusted hosts in the cloud. Since the proposed approach is independent of a third party, it is less prone to attack as it reduces the risk of correlation attacks and side channel attacks, but it is prone to denial of service as active bundle may also be not executed at all in the remote host.

Takabi et al., have proposed a security framework for cloud computing consisting of different modules to handle security and trust issues of key components [61]. The main issues discussed in the paper are identity management, access control, policy integration among multiple clouds, trust management between different clouds and between cloud providers and users. The framework identifies three main players in the cloud. They are cloud customers, service integrators and service providers. The service integrator plays the role of the mediator who brings the customers and service providers together. Service integrator facilitates collaboration among different service providers by composing services to meet the customer requirements. It is the responsibility of the service integrator to establish and maintain trust between provider domains and providers and customers. The service integrator discover the services from service providers or other service integrators, negotiate and integrate services to form collaborating services that will be sold to customers.

The service integrator module is composed of security management module, trust management module, service management module and heterogeneity management module. The heterogeneity management module manages the heterogeneity among the service providers. In addition to the above modules there are other minor modules that handle small but important tasks.

In overall this is a very comprehensive framework. But the authors have not discussed the interoperability issue of each component in the framework or implemented a prototype to evaluate the function and efficiency of the components or the overall framework.

Table 1 summarizes the proposed cloud computing trust management systems under different cloud computing parameters. From this table it is evident that most of the models proposed remain short of implementation and only a few have been simulated to prove the concept. Also, there is no single model that meets all the requirements of a cloud architecture especially the identity management, security of both data and applications, heterogeneity and SLA management. Also none of these systems have been based on solid theoretical foundation such as the trust models have been discussed in Section IV.

TABLE I
SUMMARY AND COMPARISON OF CLOUD COMPUTING TRUST MANAGEMENT SYSTEMS
SUPPORT ACROSS MULTIPLE HETEROGENEOUS CLOUDS

| Work | Type | Identity Mgmt/ Authentication | Data Security | Cloud Layer | SLA Support | Heterogeneity Support* | Implemented | Comments |
|---|---|---|---|---|---|---|---|---|
| [47] | - | Discussed | Discussed | - | - | Yes | No | No concrete proposal. Only discussed the issues. |
| [48] | Complete Platform | No | Yes | SaaS | No | Yes | No | Only a mechanism has been proposed. No implementation or evaluation carried out. |
| [49] | Social security based | Discussed | Discussed | - | Discussed | No | No | No concrete proposal. Only discussed the issues. |
| [50] | Domain based | No | No | SaaS PaaS IaaS | No | Yes | No | Model has been tested using simulation. |
| [51 - 52] | Family gene based | Discussed | No | - | No | No | No | Model has been tested using simulation. |
| [53] | Integrated with CARE Resource Broker | Yes | Yes | - | No | Yes | No | Model has been tested using simulation. |
| [54 - 55] | Built on trusted platform service | Yes | Yes | IaaS | No | Yes | No | Only a model has been proposed. |
| [56] | - | No | No | - | Yes | Yes | No | Only a model has been proposed. |
| [57] | Built on Trusted Computing Platform | No | No | IaaS | No | No | Prototype Implemented | Concept has been proved with a prototype. |
| [58] | Domain based | No | No | - | No | Yes | No | Model has been tested using simulation. |
| [59] | Watermark based security | No | No | SaaS | No | No | Prototype Implemented | Concept has been proved with a prototype. |
| [60] | Based on active bundles scheme | Yes | No | - | No | Yes | Prototype Implemented | Concept has been proved with a prototype. |
| [61] | - | Yes | No | SaaS PaaS IaaS | No | Yes | No | Only a model has been proposed |

VI. CONCLUSIONS

Cloud computing has been the new paradigm in distributed computing in the recent times. For cloud computing to become widely adopted several issues need to be addressed. Cloud security is one of the most important issues that has to be addressed. Trust management is one of the important component in cloud security as cloud environment will have different kinds of users, providers and intermediaries. Proper trust management will help the users select the provider based on their requirements and trust worthiness. Also, trust management would help the providers select the clients who are trustworthy to serve.

In the paper, a comprehensive survey has been carried out on the trust management systems implemented on distributed systems with a special emphasis cloud computing. There are several trust models proposed for distributed systems. These models were mainly proposed for systems like clusters, grids and wireless sensor networks. These models have not been used or tested in cloud computing environments. Hence the suitability of these models for use in cloud computing cannot be recommended without an extensive evaluation. The authors propose to evaluate these models in future work. The trust management systems proposed for cloud computing have been extensively studied with respect to their capability, their applicability in practical heterogonous cloud environment and their implementabilty. The results have been presented in table for easy reference. During the evaluation of these systems, it was found that none of the proposed systems is based on solid theoretical foundation and also does not take any quality of service attribute for forming the trust scores. Hence solid theoretical foundation for building trust systems for cloud computing is necessary. The theoretical basis required can be achieved by adapting the trust models proposed for other distributed systems.

REFERENCES

[1] Sheikh Mahbub Habib, Sebastian Ries, and Max Mühlhäuser, "Cloud Computing Landscape and Research Challenges regarding Trust and Reputation," in *Symposia and Workshops on Ubiquitous, Autonomic and Trusted Computing*, Xi'an, China, 2010, pp. 410-415.

[2] Rajkumar Buyya, Chee Shin Yeo, Srikumar Venugopal, James Broberg, and Ivona Brandic, "Cloud computing and emerging IT platforms: Vision, hype, and reality for delivering computing as the 5th utility," *Journal of Future Generation Computer Systems*, vol. 25, no. 6, pp. 599-616, June 2009.

[3] Radu Prodan and Simon Ostermann, "A Survey and Taxonomy of Infrastructure as a Service and Web Hosting Cloud Providers," in *10th IEEE/ACM International Conference on Grid Computing*, Banff, AB, Canada, 2009, pp. 17-25.

[4] Christian Vecchiola, Suraj Pandey, and Rajkumar Buyya, "High-Performance Cloud Computing: A View of Scientific Applications," in *10th International Symposium on Pervasive Systems, Algorithms, and Networks (ISPAN)*, Kaohsiung, Taiwan, 2009, pp. 4-16.

[5] Michael Boniface et al., "Platform-as-a-Service Architecture for Real-Time Quality of Service Management in Clouds," in *Fifth International Conference on Internet and Web Applications and Services (ICIW)*, Barcelona, Spain, 2010, pp. 155-160.

[6] Han Yu, Zhiqi Shen, Chunyan Miao, Cyril Leung, and Dusit Niyato, "A Survey of Trust and Reputation Management Systems in Wireless Communications," *Proceedings of the IEEE*, vol. 98, no. 10, pp. 1755-1772, October 2010.

[7] Zaobin Gan, Juxia He, and Qian Ding, "Trust relationship modelling in e-commerce-based social network," in *International conference on computational intelligence and security*, Beijing, China, 2009, pp. 206-210.

[8] D Harrison McKnight and Norman L Chervany, "Conceptualizing Trust: A Typology and E-Commerce Customer Relationships Model," in *34th Hawaii International Conference on System Sciences*, Island of Maui, HI, USA, 2001.

[9] Wei Wang and Guo Sun Zeng, "Bayesian cognitive trust model based self-clustering algorithm for MANETs," *Science China Information Sciences*, vol. 53, no. 3, pp. 494–505, 2010.

[10] Mario Gómez, Javier Carbó, and Earle Clara Benac, "A cognitive trust and reputation model for the ART testbed," *Inteligencia Artificial. Revista Iberoamericana de Inteligencia Artificial (in English)*, vol. 12, no. 39, pp. 29-40, 2008.

[11] Huangmao Quan and Jie Wu, "CATM: A cognitive-inspired agent-centric trust model for online social networks," in *Ninth Annual IEEE International Conference on Pervasive Computing and Communications (Percom)*, Seattle, WA, USA, 2011.

[12] Cristiano Castelfranchi, Rino Falcone, and Giovanni Pezzulo, "Trust in information sources as a source for trust: a fuzzy approach," in *Proceedings of the second international joint conference on autonomous agents and multiagent systems (AAMAS '03)*, Melbourne, Australia, 2003, pp. 89-96.

[13] Stefano De Paoli et al., "Toward trust as result: An interdisciplinary approach," *Proceedings of ALPIS, Sprouts: Working Papers on Information Systems*, vol. 10, no. 8, 2010.

[14] Masoud Akhoondi, Jafar Habibi, and Mohsen Sayyadi, "Towards a model for inferring trust in heterogeneous social networks," in *Second Asia International Conference on Modelling & Simulation*, Kuala Lumpur, Malaysia, 2008, pp. 52-58.

[15] Ram Alexander Menkes, "An economic analysis of trust, social capital, and the legislation of trust," Ghent, Belgium, LLM Thesis 2007.

[16] Jie Zhang and Robin Cohen, "Design of a mechanism for promoting honesty in e-marketplaces," in *22nd Conference on Artificial Intelligence (AAAI), AI and the Web Track*, Vancouver, British Columbia, Canada, 2007.

[17] Jie Zhang, "Promoting Honesty in Electronic Marketplaces: Combining Trust Modeling and Incentive Mechanism Design," Waterloo, Ontario, Canada, PhD Theis 2009.

[18] Shashi Mittal and Kalyanmoy Deb, "Optimal strategies of the iterated prisoner's dilemma problem for multiple conflicting objectives," in *IEEE Symposium on Computational Intelligence and Games*, Reno, NV, USA, 2006, pp. 197 - 204.

[19] Jian Zhou, Jiangbo Wang, Rongshan Liang, and Yanfu Zhang, "Flexible service analysis based on the "Prisoner's Dilemma of service"," in *6th International Conference on Service Systems and Service Management (ICSSSM '09)*, Xiamen, china, 2009, pp. 434 - 437.

[20] Hongbing Huang, Guiming Zhu, and Shiyao Jin, "Revisiting trust and reputation in multi-agent systems," in *ISECS International Colloquium on Computing, Communication, Control, and Management*, Guangzhou, China, 2008, pp. 424-429.

[21] Lik Mui, "Computational models of trust and reputation:agents, evolutionary games, and social networks," Boston, MA, USA, PhD Thesis 2002.

[22] Mohammad Momani and Subhash Challa, "Survey of Trust Models in Different Network Domains," *International Journal of Ad hoc, Sensor & Ubiquitous Computing* , vol. 1, no. 3, pp. 1-19, September 2010.

[23] Tzu Yu Chuang, "Trust with Social Network Learning in E-Commerce," in *IEEE International Conference on Communications Workshops*

*(ICC)*, Capetown, South Africa, 2010, pp. 1-6.

[24] Marcim Adamski et al., "Trust and Security in Grids: A State of the Art," European Union, 2008.

[25] Antonios Gouglidis and Ioannis Mavridis, "A Foundation for Defining Security Requirements in Grid Computing," in *13th Panhellenic Conference on Informatics ( PCI '09)*, Corfu, Greece, 2009, pp. 180-184.

[26] Leonardo B De Oliveira and Carlos A Maziero, "A Trust Model for a Group of E-mail Servers," *CLEI Electronic Journal*, vol. 11, no. 2, pp. 1-11, 2008.

[27] Qing Zhang, Ting Yu, and Keith Irwin, "A Classification Scheme for Trust Functions in Reputation-Based Trust Management," in *International Workshop on Trust, Security, and Reputation on the Semantic Web*, Hiroshima, Japan, 2004.

[28] Ruichuan Chen, Xuan Zhao, Liyong Tang, Jianbin Hu, and Zhong Chen, "CuboidTrust: A Global Reputation-Based Trust Model in Peer-to-Peer Networks," in *Autonomic and Trusted Computing*. Berlin / Heidelberg: Springer, 2007, vol. 4610, pp. 203-215.

[29] Sepandar D Kamvar, Mario T Schlosser, and Hector Garcia-Molina, "The EigenTrust Algorithm for Reputation Management in P2P Networks," in *Proceedings of the 12th international conference on World Wide Web (WWW '03)*, Budapest, Hungary, 2003, pp. 640-651.

[30] Yong Wang, Vinny Cahill, Elizabeth Gray, Colin Harris, and Lejian Liao, "Bayesian network based trust management," in *Autonomic and Trusted Computing*. Berlin / Heidelberg: Springer, 2006, pp. 246-257.

[31] Huirong Tian, Shihong Zou, Wendong Wang, and Shiduan Cheng, "A Group Based Reputation System for P2P Networks," in *Autonomic and Trusted Computing*. Berlin / Heidelberg: Springer, 2006, pp. 342-351.

[32] Wei Wang, Guosun Zeng, and Lulai Yuan, "Ant-based Reputation Evidence Distribution in P2P Networks," in *Fifth International Conference Grid and Cooperative Computing (GCC 2006)*, Hunan, China, 2006, pp. 129 - 132.

[33] Yu Zhang, Huajun Chen, and Zhaohui Wu, "A Social Network-Based Trust Model for the Semantic Web," in *Autonomic and Trusted Computing*. Berlin / Heidelberg: Springer, 2006, pp. 183-192.

[34] Fajiang Yu, Huanguo Zhang, Fei Yan, and Song Gao, "An Improved Global Trust Value Computing Method in P2P System," in *Autonomic and Trusted Computing*. Berlin / Heidelberg: Springer, 2006, pp. 258-267.

[35] Weijie Wang, Xinsheng Wang, Shuqin Pan, and Ping Liang, "A New Global Trust Model based on Recommendation for Peer-To-Peer Network," in *International Conference on New Trends in Information and Service Science*, Beijing, China, 2009, pp. 325-328.

[36] Xueming Li and Jianke Wang, "A Global Trust Model of P2P Network Based on Distance-Weighted recommendation," in *IEEE International Conference on Networking, Architecture, and Storage*, Hunan, China, 2009, pp. 281-284.

[37] Xiong Li and Liu Ling, "PeerTrust: supporting reputation-based trust for peer-to-peer electronic communities," *IEEE Transactions on Knowledge and Data Engineering*, vol. 16, no. 7, pp. 843-857, July 2004.

[38] Ayman Tajeddine, Ayman Kayssi, Ali Chehab, and Hassan Artail, "PATROL-F - A Comprehensive Reputation-Based Trust Model with Fuzzy Subsystems," in *Autonomic and Trusted Computing*. Berlin / Heidelberg: Springer, 2006, pp. 205-216.

[39] Yuan Wang, Ye Tao, Ping Yu, Feng Xu, and Jian Lü, "A Trust Evolution Model for P2P Networks," in *Autonomic and Trusted Computing*. Berlin / Heidelberg: Springer, 2007, pp. 216-225.

[40] Zhuo Tang, Zhengding Lu, and Kai Li, "Time-based Dynamic Trust Model using Ant Colony Algorithm," *Wuhan University Journal of Natural Sciences*, vol. 11, no. 6, pp. 1462-1466, 2006.

[41] Felix Gomez Marmol, Gregorio Martinez Perez, and Antonio F Gomez Skarmeta, "TACS, a Trust Model for P2P Networks," *Wireless Personal Communications*, vol. 51, no. 1, pp. 153-164, 2009.

[42] Ghada Derbas, Ayman Kayssi, Hassan Artail, and Ali Chehab, "TRUMMAR - A Trust Model for Mobile Agent Systems based on Reputation," in *The IEEE/ACS International Conference on Pervasive Services (ICPS 2004)*, Beirut, Lebanon, 2004, pp. 113-120.

[43] Ayman Tajeddine, Ayman Kayssi, Ali Chehab, and Hassan Artail, "PATROL: A Comprehensive Reputation-based Trust Model," *International Journal of Internet Technology and Secured Transactions*, vol. 1, no. 1/2, pp. 108-131, August 2007.

[44] Felix Gomez Marmol, Gregorio Mrtinez Perez, and Javier G Marin-Blazquez, "META-TACS: A Trust Model Demonstration of Robustness through a Genetic Algorithm," *Autosof Journal of Intelligent Automation and Soft Computing*, vol. 16, no. X, pp. 1-19, 2009.

[45] Cesar Ghali, Ali Chehab, and Ayman Kayssi, "CATRAC: Context-Aware Trust- and Role-based Access Control for Composite Web Services," in *10th IEEE International Conference on Computer and Information Technology*, Bradford, England, 2010, pp. 1085-1089.

[46] Yao Wang and Julita Vassileva, "Bayesian Network-based Trust Model," in *IEEE/WIC International Conference on Web Intelligence (WI 2003)*, Halifax, Canada, 2003, pp. 372 - 378.

[47] Khaled M Khan and Qutaibah Malluhi, "Establishing Trust in Cloud Computing," *IT Professional*, vol. 12, no. 5, pp. 20 - 27, 2010.

[48] Zhexuan Song, Jusus Molina, and Christina Strong, "Trusted Anonymous Execution: A Model to RaiseTrust in Cloud," in *9th International Conference on Grid and Cooperative Computing (GCC)*, Nanjing, China, 2010, pp. 133 - 138.

[49] Hiroyuki Sato, Atsushi Kanai, and Shigeaki Tanimoto, "A Cloud Trust Model in a Security Aware Cloud," in *10th IEEE/IPSJ International Symposium on Applications and the Internet (SAINT)*, Seoul, South Korea, 2010, pp. 121 - 124.

[50] Wenjuan Li, Lingdi Ping, and Xuezeng Pan, "Use trust management module to achieve effective security mechanisms in cloud environment," in *International Conference on Electronics and Information Engineering (ICEIE)*, vol. 1, Kyoto, Japan, 2010, pp. 14-19.

[51] Tie Fang Wang, Bao Sheng Ye, Yun Wen Li, and Yi Yang, "Family Gene based Cloud Trust Model," in *International Conference on Educational and Network Technology (ICENT)*, Qinhuangdao, China, 2010, pp. 540 - 544.

[52] Tie Fang Wang, Bao Sheng Ye, Yun Wen Li, and Li Shang Zhu, "Study on Enhancing Performance of Cloud Trust Model with Family Gene Technology," in *3rd IEEE International Conference on Computer Science and Information Technology (ICCSIT)*, vol. 9, Chengdu, China, 2010, pp. 122 - 126.

[53] Paul D Manuel, Thamarai Selve, and Mostafa Ibrahim Abd-EI Barr, "Trust management system for grid and cloudresources," in *First International Conference on Advanced Computing (ICAC 2009)*, Chennai, India, 2009, pp. 176-181.

[54] Zhidong Shen, Li Li, Fei Yan, and Xiaoping Wu, "Cloud Computing System Based on TrustedComputing Platform," in *International Conference on Intelligent Computation Technology and Automation (ICICTA)*, vol. 1, Changsha, China, 2010, pp. 942 - 945.

[55] Zhidong Shen and Qiang Tong, "The security of cloud computing system enabled by trusted computing technology," in *2nd International Conference on Signal Processing Systems (ICSPS)*, vol. 2, Dalian, China, 2010, pp. 11-15.

[56] Mohammed Alhamad, Tharam Dillon, and Elizabeth Chang, "SLA-based Trust Model for Cloud Computing," in *13th International Conference on Network-Based Information Systems*, Takayama, Japan, 2010, pp. 321 - 324.

[57] Xiao Yong Li, Li Tao Zhou, Yong Shi, and Yu Guo, "A trusted computing environment model in cloudarchitecture," in *Ninth International Conference on Machine Learning and Cybernetics (ICMLC)*, vol. 6, Qingdao, China, 2010, pp. 2843-2848.

[58] Zhimin Yang, Lixiang Qiao, Chang Liu, Chi Yang, and Guangming Wan, "A Collaborative Trust Model of Firewall-through based on Cloud

Computing," in *14th International Conference on Computer Supported Cooperative Work in Design (CSCWD)*, Shanghai, China, 2010, pp. 329 - 334.

[59] Junning Fu, Chaokun Wang, Zhiwei Yu, Jianmin Wang, and Jia Guang Sun, "A Watermark-Aware Trusted Running Environment for Software Clouds," in *Fifth Annual ChinaGrid Conference (ChinaGrid)*, Guangzhou, China, 2010, pp. 144 - 151.

[60] Rohit Ranchal et al., "Protection of Identity Information in Cloud Computing without Trusted Third Party," in *29th IEEE International Symposium on Reliable Distributed Systems*, New Delhi, India, 2010, pp. 1060-9857.

[61] Hassan Takabi, James B.D Joshi, and Gail Joon Ahn, "SecureCloud: Towards a Comprehensive Security Framework for Cloud Computing Environments," in *34th Annual IEEE Computer Software and Applications Conference Workshops*, Seoul, South Korea, 2010, pp. 393 - 398.